\documentclass[onecolumn,prl]{revtex4}
\usepackage{bm}
\usepackage{epsfig}
\usepackage{amsmath}
\begin{document}
\title{Some exact results for the Smoluchowski equation for a parabolic potential with time dependent delta function sink.
}
\author{Diwaker and Aniruddha Chakraborty \\
School of Basic Sciences, Indian Institute of Technology Mandi,\\
Mandi, Himachal Pradesh, 175001, India.}
\date{\today }
\begin{abstract}
\noindent The Smoluchowski equation with a time dependent delta function sink is solved exactly for many special cases. In all other cases the problem can be reduced to an integral equation. It is shown that by knowing the probability distribution at the position of sink, one can derive analytical expression for probability distribution everywhere. Thus the problem is reduced from a PDE in two variables to an integral equation of one. As far as the authors knowledge, we are the first one to provide an exact analytical solution of Smoluchowski equation for a parabolic potential with time dependent sink.  
\end{abstract}
\maketitle
\noindent 
Diffusion-controlled reactions have been classical topics \cite{Zhang}. There is huge interest in this subject because of it's wider connections recognized in physics, chemistry and biology \cite{Weiss}. One particular model of this class is a diffusion under a random-trapping environment. This problem can be modeled by a particle undergoing diffusive motion in presence of a time dependent Delta function sink. One may describe the diffusion motion by Smoluchowski equation and the sink could be a Dirac Delta function of arbitrary position and strength. As per the authors knowledge this Letter is the first one to consider the exact analytical solution of Smoluchowski equation with time dependent sink. All the solutions available in literature are the cases where the strength of the Dirac delta function used to model the sink is assumed to be time independent \cite{diw1,diw2,diw3,diw4,diw5,diw6}, although the sink has to be time dependent in general in most of the cases and but the way strength of the sink depends on time is not well understood. In contrast, in the present work we have considered the cases where the strength of the sink varies with time. There are two common approaches one can think of using for solving this type of problems. One is based on using path integral method of Feynman type and the other is based on Laplace transforms. We use the latter method, although both are closely related. In the following we would like to solve the Smoluchowski equation with a time dependent sink.
\begin{equation}
\frac{\partial}{\partial t}P(x,t) = \left[L - k(t)\delta(x)\right]P(x,t).
\end{equation}
In the above we denote that the probability of the particle to survive on the potential energy surface as $P(x,t)$, and 
\begin{equation}
L = A\frac{\partial^2}{\partial x^2}+\frac{A}{k_{b}T}\frac{\partial}{\partial x}\frac{dV(x)}{dx}.\
\end{equation}
The schematic view of our problem is given in Fig. 1.
\begin{figure}
\centering
{\includegraphics[width=110mm]{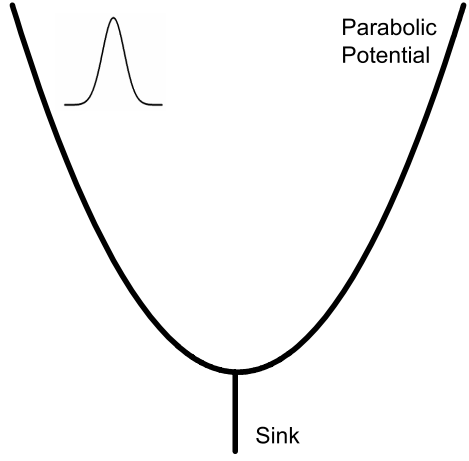}}
\caption{
Schematic diagram showing the formulation of our problem}
\end{figure}
We define the Laplace transform $\overline{P}(x,s) = L[ k(t)P(x,t)] = \int_{0}^{\infty}P(x,t)e^{- s t}dt$, then the Laplace transform of Eq. (1) can be written as
\begin{equation}
\left[s - L )\right]\overline{P}(x,s)+ L[ k(t) \delta(x)P(x,t)]  = P_{0}(x,0).
\end{equation}
We will now solve Eq. (2), however we must first consider solving the following homogeneous equation \cite{diw7} 
\begin{equation}
\left[s - L\right]\overline{P}(x,s) = P_{0}(x,0).
\end{equation}
Solution of this homogeneous equations must satisfy the boundary conditions given by 
\begin{eqnarray}
\frac{d\overline{P}(0^{+},s)}{dx} - \frac{d\overline{P}(0^{-},s)}{dx}= \frac{L}{A}[k(t)P(0,t)].
\end{eqnarray}
In case of a parabolic potential represented by V(x) = $\frac{1}{2}m\omega^2x^2$, Eq. (4) can be written as
\begin{equation}
\left[s -A\frac{\partial^2}{\partial x^2}-\frac{\partial}{\partial x}\beta x\right]\overline{P}(x,s) = P_{0}(x,0)
\end{equation}
with $\beta = \frac{m\omega^2 A}{k_{B}T} = \frac{m\omega^2}{\eta}$. Now we introduce the variable $z$ and $z_0$ by $ z = x(\frac{A}{\beta})^{\frac{1}{2}}, z_{0} = x_{0}(A/B)^{1/2}$.
The solution of above equation in terms of parabolic cylinder functions can be written as
\begin{eqnarray}
\overline{P}(z,s) = a(s)\frac{D_{v}(-z_{<})D_{v}(-z_{>})exp[-z^2/4]\Gamma(1-v)[\frac{B}{2\pi A}]^{\frac{1}{2}}}{s}\nonumber \\+\int_{-\infty}^{\infty}\frac{D^{*}_{v}(-z_{<})D^{*}_{v}(-z_{>})exp[(z_{0}^2-z^{2})/4] \Gamma(1-v)[\frac{B}{2\pi A}]^{\frac{1}{2}}}{s}P(z_{0})dz_{0},
\end{eqnarray}
where $D_{v}$ is parabolic cylinder functions with $z_{<} = min(z,0), z_{>}= max(z,0)$ and $D^{*}_v$ is also the same parabolic cylinder function with  $z_{<} = min(z,z_0), z_{>}= max(z,z_0)$. In Eq. (7), $a(s)$ is an arbitrary function of $s$, yet to be determined. 
\newline 
\par
\noindent Now we will first evaluate the integration constant $a(s)$ for the case where $k(t) = k_0$ by using Eq. (5) as follows
\begin{eqnarray}
\frac{d\overline{P}(0^{+},s)}{dx} - \frac{d\overline{P}(0^{-},s)}{dx} = -a(s) = \frac{k_0}{A} \overline{P}(0,s).
\end{eqnarray}
Putting this value of $a(s)$ in Eq. (7) we will get 
\begin{eqnarray}
\overline{P}(z,s) = - \frac{k_0}{A} \overline{P}(0,s) \frac{D_{v}(-z_{<})D_{v}(-z_{>})\exp[-z^2/4]\Gamma(1-v)[\frac{B}{2\pi A}]^{\frac{1}{2}}}{s} \nonumber \\+\int_{-\infty}^{\infty}\frac{D^{*}_{v}(-z_{<})D^{*}_{v}(-z_{>}) exp[(z_{0}^2-z^{2})/4]\Gamma(1-v)[\frac{B}{2\pi A}]^{\frac{1}{2}}}{s}P(z_{0})dz_{0}.
\end{eqnarray}
We now set $z = 0$ in the above equation to get
\begin{equation}
\overline{P}(0,s) = - \frac{k_0}{A} \overline{P}(0,s) \frac{D_{v}(0)D_{v}(0)\Gamma(1-v)[\frac{B}{2\pi A}]^{\frac{1}{2}}}{s}+\int_{-\infty}^{\infty}\frac{D^{*}_{v}(0)D^{*}_{v}(0)exp[z_{0}^2/4]\Gamma(1-v)[\frac{B}{2\pi A}]^{\frac{1}{2}}}{s}P(z_{0})dz_{0}.
\end{equation}
Now we solve the above equation for $\overline{P}(0,s)$ to get
\begin{equation}
\overline{P}(0,s) = \frac{\int_{-\infty}^{\infty}\frac{D^{*}_{v}(0)D^{*}_{v}(0)\exp[z_{0}^2/4]\Gamma(1-v)[\frac{B}{2\pi A}]^{\frac{1}{2}}}{s}P(z_{0})dz_{0}}{1+ \frac{k_0}{A} \frac{D_{v}(0)D_{v}(0)\Gamma(1-v)[\frac{B}{2\pi A}]^{\frac{1}{2}}}{s}}.
\end{equation}
This, when substituted back into Eq. (9) gives
\begin{eqnarray}
\overline{P}(z,s) = - \frac{k_0}{A} \frac{D_{v}(0)D_{v}(0)\Gamma(1-v)[\frac{B}{2\pi A}]^{\frac{1}{2}}}{s} \frac{\int_{-\infty}^{\infty}\frac{D^{*}_{v}(0)D^{*}_{v}(0)\exp[z_{0}^2/4]\Gamma(1-v)[\frac{B}{2\pi A}]^{\frac{1}{2}}}{s}P(z_{0})dz_{0}}{1+ \frac{k_0}{A} \frac{D_{v}(0)D_{v}(0)\Gamma(1-v)[\frac{B}{2\pi A}]^{\frac{1}{2}}}{s}}\nonumber \\ +\int_{-\infty}^{\infty}\frac{D^{*}_{v}(-z_{<})D^{*}_{v}(-z_{>}) \exp[(z_{0}^2-z^{2})/4]\Gamma(1-v)[\frac{B}{2\pi A}]^{\frac{1}{2}}}{s}P(z_{0})dz_{0}.
\end{eqnarray}
Also in some situations it is always advantageous to present the solution in the following form 
\begin{equation}
\overline{P}(z,s) = \int_{-\infty}^{\infty}G(z,s|z_{0})P(z_{0})dz_{0},
\end{equation}
where $G(z,s|z_{0})$ is nothing but Greens function. By comparing Eq. (10) and Eq. (11) we find 
\begin{eqnarray}
G(z,s|z_{0}) = - \frac{k_0}{A} \frac{D_{v}(0)D_{v}(0)\Gamma(1-v)[\frac{B}{2\pi A}]^{\frac{1}{2}}}{s} \frac{\frac{D^{*}_{v}(0)D^{*}_{v}(0)exp[z_{0}^2/4]\Gamma(1-v)[\frac{B}{2\pi A}]^{\frac{1}{2}}}{s}}{1+ \frac{k_0}{A} \frac{D^{*}_{v}(0)D^{*}_{v}(0)\Gamma(1-v)[\frac{B}{2\pi A}]^{\frac{1}{2}}}{s}}\nonumber \\+\frac{D^{*}_{v}(-z_{<})D^{*}_{v}(-z_{>}) \exp[(z_{0}^2-z^{2})/4]\Gamma(1-v)[\frac{B}{2\pi A}]^{\frac{1}{2}}}{s}.
\end{eqnarray}
Our expression for $G(z,s|z_{0})$ as given above is the same as the one derived by Sebastian using another method \cite{diw6}.
\newline
\par
\noindent Now we consider the case, where  $k(t) =  - \alpha t$, so that an equation similar to Eq. (8) would be
\begin{eqnarray}
\frac{d\overline{P}(0^{+},s)}{dx} - \frac{d\overline{P}(0^{-},s)}{dx} = -a(s) = \frac{L}{A}[k(t)P(0,t)] = \int_{0}^{\infty}e^{-st}-\frac\alpha t P(0,t)dt \nonumber \\ = \frac{\alpha}{A} \frac{\partial}{\partial s}\overline{P}(0,s)\;;\; a(s) = -\frac{\alpha}{A} \frac{\partial}{\partial s}\overline{P}(0,s).
\end{eqnarray}
Putting this solution for $a(s)$ in Eq. (7) we will get 
\begin{eqnarray}
\overline{P}(z,s) = -\frac{\alpha}{A} \left[\frac{\partial}{\partial s}\overline{P}(0,s)\right]\frac{D_{v}(-z_{<})D_{v}(-z_{>})\exp[-z^2/4]\Gamma(1-v)[\frac{B}{2\pi A}]^{\frac{1}{2}}}{s} \nonumber \\+\int_{-\infty}^{\infty}\frac{D^{*}_{v}(-z_{<})D^{*}_{v}(-z_{>})\exp[(z_{0}^2-z^{2})/4]\Gamma(1-v)[\frac{B}{2\pi A}]^{\frac{1}{2}}}{s}P(z_{0})dz_{0}.
\end{eqnarray}
Now we consider $P(z_{0}) = \delta(z'_0 -z_{0})$ as the initial condition, then above equation simplifies considerably to the following one (after we replace $z_{0}'$ by $z_0$) 
\begin{eqnarray}
\overline{P}(z,s) =-\frac{\alpha}{A} \left[\frac{\partial}{\partial s}\overline{P}(0,s)\right]\frac{D_{v}(-z_{<})D_{v}(-z_{>})\exp[-z^2/4]\Gamma(1-v)[\frac{B}{2\pi A}]^{\frac{1}{2}}}{s} \nonumber \\+\frac{D^{*}_{v}(-z_{<})D^{*}_{v}(-z_{>})\exp[(z_{0}^2-z^{2})/4]\Gamma(1-v)[\frac{B}{2\pi A}]^{\frac{1}{2}}}{s}.
\end{eqnarray}
For $ z= 0 $ the above equation can be rewritten as
\begin{eqnarray}
\overline{P}(0,s) =-\frac{\alpha}{A} \left[\frac{\partial}{\partial s}\overline{P}(0,s)\right]\frac{D_{v}(0)D_{v}(0)\Gamma(1-v)[\frac{B}{2\pi A}]^{\frac{1}{2}}}{s} \nonumber \\+\frac{D^{*}_{v}(0)D^{*}_{v}(0)\exp[(z_{0}^2)/4]\Gamma(1-v)[\frac{B}{2\pi A}]^{\frac{1}{2}}}{s}.
\end{eqnarray}
Dividing the above equation on both sides by $ \frac{D_{v}(0)D_{v}(0)}{s}\Gamma(1-v)[\frac{B}{2\pi A}]^{\frac{1}{2}}$, we get
\begin{eqnarray}
\frac{s}{D_{v}(0)D_{v}(0)\Gamma(1-v)[\frac{B}{2\pi A}]^{\frac{1}{2}}}\overline{P}(0,s) = -\frac{\alpha}{A}\frac{\partial}{\partial s}\overline{P}(0,s)+\frac{D^{*}_{v}(0)D^{*}_{v}(0)\exp[z_{0}^{2}/4]}{D_{v}(0)D_{v}(0)}.
\end{eqnarray}
Let us consider the following substitutions, we put 
\begin{eqnarray}
\beta(s) = \frac{s}{D_{v}(0)D_{v}(0)\Gamma(1-v)[\frac{B}{2\pi A}]^{\frac{1}{2}}}  \;\;\;\; {\text and} \\
\gamma(s) = \frac{D^{*}_{v}(0)D^{*}_{v}(0)\exp[z_{0}^{2}/4]}{D_{v}(0)D_{v}(0)}.
\end{eqnarray}
Hence Eq. (19) can be rewritten as
\begin{eqnarray}
\frac{\partial}{\partial s}\overline{P}(0,s) + \frac{A \beta(s)}{\alpha} \overline{P}(0,s) = \frac{A \gamma(s) }{\alpha}.
\end{eqnarray}
Multiplying the above equation on both sides by $e^{f(s)}$, where $ \frac{df(s)}{ds} = \frac{A \beta (s)}{\alpha}$ and using some simple mathematics we reach at finally
\begin{equation}
\overline{P}(0,s) = \frac{A e^{-f(s)}}{\alpha}\int_{s}^{\infty} \gamma(s') ds',
\end{equation}
where $f(s) = \frac{A}{\alpha}\int_{0}^{s} \beta(s') ds'$. Hence we have found exact analytical solution for $\overline{P}$(0,\;s) and once it is known we are able to determine the probability distribution function $\overline{P}(z, s)$ everywhere using Eq. (16).
\newline
\par
\noindent
Now we consider the case, where  $k(t) =  - \frac{\alpha}{t}$, so that an equation similar to Eq. (8) would be
\begin{eqnarray}
\frac{d\overline{P}(0^{+},s)}{dx} - \frac{d\overline{P}(0^{-},s)}{dx} = - a(s) = \frac{L}{A}[k(t)P(0,t)] = \int_{0}^{\infty}e^{-st}\frac{\alpha}{t}P(0,t)dt \nonumber \\ = -\frac{\alpha}{A}\int_{s}^{\infty}\overline{P}(0,s')ds^{'}\;;\; a(s) = \frac{\alpha}{A}\int_{s}^{\infty}\overline{P}(0,s')ds^{'}.
\end{eqnarray}
Putting this value of $a(s)$ in Eq. (7) we  get 
\begin{eqnarray}
\overline{P}(z,s) = \frac{\alpha}{A}\left[\int_{s}^{\infty}\overline{P}(0,s')ds^{'}\right]\frac{D_{v}(-z_{<})D_{v}(-z_{>})\exp[-z^2/4]\Gamma(1-v)[\frac{B}{2\pi A}]^{\frac{1}{2}}}{s} \nonumber \\+\int_{-\infty}^{\infty}\frac{D^{*}_{v}(-z_{<})D^{*}_{v}(-z_{>})\exp[(z_{0}^2-z^{2})/4]\Gamma(1-v)[\frac{B}{2\pi A}]^{\frac{1}{2}}}{s}P(z_{0})dz_{0}.
\end{eqnarray}
Now we consider $P(z_{0}) = \delta(z'_0 -z_{0})$ as the initial condition, then above equation simplifies considerably to the following one (after we replace $z_{0}'$ by $z_0$) 
\begin{eqnarray}
\overline{P}(z,s) = \frac{\alpha}{A}\left[\int_{s}^{\infty}\overline{P}(0,s')ds^{'}\right]\frac{D_{v}(-z_{<})D_{v}(-z_{>})\exp[-z^2/4]\Gamma(1-v)[\frac{B}{2\pi A}]^{\frac{1}{2}}}{s} \nonumber \\+\frac{D^{*}_{v}(-z_{<})D^{*}_{v}(-z_{>})\exp[(z_{0}^2-z^{2})/4]\Gamma(1-v)[\frac{B}{2\pi A}]^{\frac{1}{2}}}{s}.
\end{eqnarray}
For z=$0$ the above equation can be rewritten as
\begin{eqnarray}
\overline{P}(0,s) = \frac{\alpha}{A}\left[\int_{s}^{\infty}\overline{P}(0,s')ds^{'}\right]\frac{D_{v}(0)D_{v}(0)\Gamma(1-v)[\frac{B}{2\pi A}]^{\frac{1}{2}}}{s} \nonumber \\+\frac{D^{*}_{v}(0)D^{*}_{v}(0)\exp[z_{0}^2/4]\Gamma(1-v)[\frac{B}{2\pi A}]^{\frac{1}{2}}}{s}.
\end{eqnarray}
Dividing the above equation on both sides by $ \frac{D_{v}(0)D_{v}(0)}{s}\Gamma(1-v)[\frac{B}{2\pi A}]^{\frac{1}{2}}$, we get
\begin{eqnarray}
\frac{s}{D_{v}(0)D_{v}(0)\Gamma(1-v)[\frac{B}{2\pi A}]^{\frac{1}{2}}}\overline{P}(0,s) = \frac{\alpha}{A}\int_{s}^{\infty}\overline{P}(0,s')ds^{'}+\frac{D^{*}_{v}(0)D^{*}_{v}(0)\exp[z_{0}^{2}/4]}{D_{v}(0)D_{v}(0)}.
\end{eqnarray}
Hence the above equation can be rewritten as
\begin{eqnarray}
\beta(s) \overline{P}(0,s) = \frac{\alpha}{A}\int_{s}^{\infty}\overline{P}(0,s')ds^{'}+ \gamma (s).
\end{eqnarray}
In solving the above equation, we put $u(s) = \int_{s}^{\infty}\overline{P}(0,s')ds^{'}$, then $\frac{\partial u(s)}{\partial s} = \overline{P}(0,s)$, hence above equation can be further written as
\begin{equation}
\frac{\partial u(s)}{\partial s} - \frac{\alpha}{A \beta(s)}u(s)=\frac{\gamma(s)}{\beta(s)}.
\end{equation}
Multiplying both side of the above equation by $e^{ f(s)}$, where $ f'(s) = - \frac{\alpha}{A \beta (s)}$ and using some simple mathematics we reach at finally
\begin{equation}
u(s)= e^{-f(s)}\int_{s}^{\infty}\frac{\gamma(s')}{\beta(s')}ds'.
\end{equation}
which can be used to evaluate $\overline{P}(0,s)$, as given below
\begin{equation}
\overline{P}(0,s) = \frac{d}{ds}\left[ e^{-f(s)}\int_{s}^{\infty}\frac{\gamma(s')}{\beta(s')}ds'\right],
\end{equation}
where $f(s) = -\frac{\alpha}{A}\int_{0}^{s} \frac{ds}{\beta(s)}$. Hence we have found exact analytical solution for $\overline{P}$(0, s) and once it is known we are able to determine the probability distribution function $\overline{P}$(z, s) everywhere using Eq. (26).
\newline
\par
\noindent
Now we consider the case, where  $k(t) =\beta exp(-\alpha t)$, so that an equation similar to Eq. (8) would be
\begin{eqnarray}
\frac{d\overline{P}(0^{+},s)}{dx} - \frac{d\overline{P}(0^{-},s)}{dx} = -a(s) = \frac{L}{A}[k(t)P(0,t)] = \int_{0}^{\infty}e^{-st}\beta exp(-\alpha t)P(0,t)dt \nonumber \\ = \frac{\beta}{A} \overline{P}(0,s+\alpha)\;;\; a(s) = -\frac{\beta}{A} \overline{P}(0,s+\alpha).
\end{eqnarray}
Putting this value of $a(s)$ in Eq. (7) we will get 
\begin{eqnarray}
\overline{P}(z,s) = -\frac{\beta}{A} \overline{P}(0,s+\alpha)\frac{D_{v}(-z_{<})D_{v}(-z_{>})\exp[-z^2/4]\Gamma(1-v)[\frac{B}{2\pi A}]^{\frac{1}{2}}}{s} \nonumber \\+\int_{-\infty}^{\infty}\frac{D^{*}_{v}(-z_{<})D^{*}_{v}(-z_{>})\exp[(z_{0}^2-z^{2}/)4]\Gamma(1-v)[\frac{B}{2\pi A}]^{\frac{1}{2}}}{s}P(z_{0})dz_{0}.
\end{eqnarray}
Now we consider $P(z_{0}) = \delta(z'_0 -z_{0})$ as the initial condition, then above equation simplifies considerably to the following one (after we replace $z_{0}'$ by $z_0$) 
\begin{eqnarray}
\overline{P}(z,s) = -\frac{\beta}{A} \overline{P}(0,s+\alpha)D_{v}(-z_{<})D_{v}(-z_{>})\exp[-z^2/4]\Gamma(1-v)[\frac{B}{2\pi A}]^{\frac{1}{2}} \nonumber \\+D^{*}_{v}(-z_{<})D^{*}_{v}(-z_{>})\exp[(z_{0}^2-z^{2})/4]\Gamma(1-v)[\frac{B}{2\pi A}]^{\frac{1}{2}}.
\end{eqnarray}
For z=$0$ the above equation can be rewritten as
\begin{eqnarray}
\overline{P}(0,s) = -\frac{\beta}{A} \overline{P}(0,s+\alpha)D_{v}(0)D_{v}(0)\Gamma(1-v)[\frac{B}{2\pi A}]^{\frac{1}{2}} \nonumber \\+D^{*}_{v}(0)D^{*}_{v}(0)\exp[z_{0}^2/4]\Gamma(1-v)[\frac{B}{2\pi A}]^{\frac{1}{2}}.
\end{eqnarray}
Further to obtain a solution of the above equation we could iterate the expression repeatedly to obtain the series solution, given by 
\begin{equation}
\overline{P}(0,s) = \sum_{n = 0}^{\infty}\eta^{n}a_{n}(s).
\end{equation}
Since we already have the form of this series, we may substitute Eq. (37) directly to Eq. (36) and then solving for a$^{'}$(s) by equating like powers of $\eta$, we find that
\begin{eqnarray}
a_{n}(s) = -\frac{D_{v}(0)D_{v}(0)\exp[z_{0}^{2}/4]\Gamma(1-v)[\frac{B}{2\pi A}]^{\frac{1}{2}}}{As} \; for \; n = 0 \nonumber \\
a_{n}(s) = -\left[\frac{D_{v}(0)D_{v}(0)\exp[z_{0}^{2}/4]\Gamma(1-v)[\frac{B}{2\pi A}]^{\frac{1}{2}}}{A n s}\right]\left[\frac{D^{*}_{v}(0)D^{*}_{v}(0)\Gamma(1-v)[\frac{B}{2\pi A}]^{\frac{1}{2}}}{As}\right]^{n}\; for\; n>0
\end{eqnarray}

It was shown in this letter that the Smoluchowski equation for parabolic potential with a time dependent Dirac delta function sink can be solved exactly in many special cases. Our method can be used to find exact solution to the case where the sink is represented by a non-local operator, and may be represented by  $|f > k_0 <g|$, where $f$ and $g$ are arbitrary acceptable functions. Choosing both of them to be a Gaussian would be an improvement over the localized sink model. Sink may even be represented by a linear combination of such operators.


\end{document}